# Hypersharp Resonant Capture of Anti-Neutrinos

R. S. Raghavan

*Institute of Particle, Nuclear & Astronomical Sciences and Department of Physics*
*Virginia Polytechnic Institute and State University, Blacksburg VA 24061*

Recent ideas suggest that the 18.6 keV antineutrino ($\tilde{\nu}_e$) line from 2-body decay of $^3$H in crystals is emitted with *natural* width, motionally narrowed by lattice vibrations as in recoilless emission. It can be resonantly captured in $^3$He with geometrical cross section $\sigma \sim 10^{-17}$ cm$^2$. A key technique solves a basic obstacle for achieving resonance--the chemical difference of H and He in metals. The low $\tilde{\nu}_e$ energy, the high $\sigma$ and the hypersharp sensitivity $\Delta E/E \sim 10^{-29}$ make an extraordinary tool for bench scale tests of $\tilde{\nu}_e$ $\theta_{13}$ oscillations and predicted Planck length limits on nuclear level widths in models of quantum gravity.

Two-body weak nuclear decays emit monoenergetic lines of antineutrinos ($\tilde{\nu}_e$). The two well known modes of such decay are electron capture (EC) and the reverse process of bound-state beta decay (BB)[1] in which the β-electron is captured in an atomic orbital. The question if these lines can also be emitted *recoilless* was raised immediately after the discovery of the Mössbauer effect (ME). Visscher considered the EC mode[2] in 1959 and, 25 years later, Kells and Schiffer[3], the BB mode, particularly that of tritium $^3$H (T). However, these ideas remain yet speculative because of the very stringent, unanswered experimental demands, even for the more favorable case of T. State-of-the-art hydrogen storage technology and materials now suggest a breakthrough in the T case. In a preliminary report[4], I proposed a specific approach to observe recoilless resonant capture of the 18.6 keV $\tilde{\nu}_e$ emitted in the T-BB in a $^3$He target. The key idea focuses on solving the biggest problem posed by this experiment--the different behaviors of the noble gas He absorber and the chemically bound source T in metals.

With the advantage of recoilless emission, the resonant cross section σ for capture is fundamentally determined by the spectral widths of the emitted $\tilde{\nu}_e$. The widths are, in turn determined by the broadening induced by various means, but mainly by the spin motions and the fluctuations of local dipolar fields. In ref. 4, a relaxation width measured by NMR in the chosen material was used and an effective resonance cross section σ ~3x10$^{-33}$ cm$^2$ some 10$^{10}$ times that for usual $\tilde{\nu}_e$ reactions, was derived. While this was very attractive, major experimental challenges remained.

New ideas have recently emerged (see companion paper[5]), on the origins of the linewidth. The broadening assumed in ref. 4 is appropriate for short lived states (including all ME cases so far) but not for very long lived states for which no data is available yet. Ref. 5 suggests surprisingly, that in these cases, one should actually expect hypersharp $\tilde{\nu}_e$ lines of *natural line width*--not the severe broadening assumed in ref. 4, basically because it ignores the key role of motional averaging via lattice vibrations in full analogy to recoilless emission itself. In this case σ rises to the geometrical limit, vastly larger than the previous estimate and dramatically enhances prospects for observing resonant capture of tritium $\tilde{\nu}_e$. In this Letter I briefly summarize the theory of ref.5, and on that basis, discuss a radically simplified experimental approach.

Hypersharp $\tilde{\nu}_e$ lines offer a basically new tool of unprecedented power, combining the low $\tilde{\nu}_e$ energy, the high resonance σ ~ 10$^{-17}$ cm$^2$ and the hypersharp energy sensitivity $\Delta E/E \sim 10^{-29}$. It challenges the imagination of new perspectives of the physical universe. One example is tests of $\tilde{\nu}_e$ $\theta_{13}$ oscillations with gram- not kiloton scale targets and bench-scale—not km scale baselines. Another is testing the predicted limits on nuclear level widths set by the Planck length.

The T-$^3$He system $^3$H(1/2)$^+ \leftrightarrow$ $^3$He(1/2)$^+ + \tilde{\nu}_e$; [(E($\tilde{\nu}_e$)=18.6 keV; τ ($^3$H) ~ 6x10$^8$ s; Γ~10$^{-24}$ eV) is ideal for resonant $\tilde{\nu}_e$ capture. It offers a sizable BB branching (~5.4 x10$^{-3}$)[1] to the atomic ground state of $^3$He. The initial T atom has a vacancy in the 1s shell for Bβ decay and the target $^3$He has two 1s electrons one of which can be captured. 1s EC is ideal since | Ψ$^2$|(He n =1s) (for BB decay and $\tilde{\nu}_e$ capture) is $\propto 1/n^3$ and maximal for n =1. Note that spins of both $^3$H and $^3$He are ½, thus both have zero quadrupole moments. In BB decay[1]

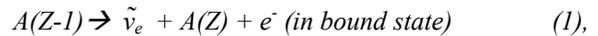

$$A(Z-1) \rightarrow \tilde{\nu}_e + A(Z) + e^- \text{ (in bound state)} \qquad (1),$$

the β- electron of A(Z-1) is inserted in a vacant orbit in A(Z). A $\tilde{\nu}_e$ line is emitted with the unique energy:

$$E_{\tilde{\nu}_e} = Q + B_Z - E_R \qquad (2)$$

where B$_z$ the shell binding energy is *gained* in inserting an electron in A(Z). E$_R$ is a deficit due to nuclear recoil. Mikaelyan et al[6] first noted the reverse reaction

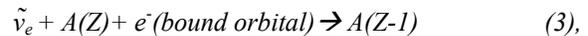

$$\tilde{\nu}_e + A(Z) + e^- \text{(bound orbital)} \rightarrow A(Z-1) \qquad (3),$$

and its resonant character at the $\tilde{\nu}_e$ energy:

$$E(\tilde{v}_{e\ res}) = Q + B_Z + E_R \qquad (4).$$

Q is the maximum $\tilde{v}_e$ energy (= $M_{Z-1} - M_Z$) in the β-decay A(Z-1)→A(Z). The $\tilde{v}_e$ from (1) has exactly the excess energy $B_z$ in (2) in order to remove the same electron in the capture (3). As is well known, the double deficit $2E_R$ remains. The basic idea of ref. 4 (and 3) is that the $\tilde{v}_e$ line from T- B decay is *resonant* if the $\tilde{v}_e$ emission and absorption are *recoilless*.

The resonance cross section σ is determined directly by the spectral density of the incident beam at the resonance energy i.e., the line width of the $\tilde{v}_e$ from the BB-decay (3)-- broader it is, less the σ. A width ~$10^{12}$ times the natural width was assumed in ref. 4. Based on ref. 5 a $\tilde{v}_e$ of *natural* width Γ should be emitted in long lived states such as T (see below). Then the spectral density at resonance is maximal, $1/\Gamma$. Thus, σ is maximal-- the geometrical value[6] $\sigma_o = 2\pi^2\lambda^2$ as in γ-ray resonance. The cross section $\sigma_o \sim 2\text{x}10^{-17}$ cm$^2$ saturates absorption in μg/cm$^2$ thick $^3$He targets. Such a "black" $\tilde{v}_e$ absorber is a novel experience in $\tilde{v}_e$ research.

Time-dependent fields such as dipolar interactions ($\propto 1/r^3$ the interatomic distance) create a fluctuating field from nuclear motions because of lattice vibrations. (the mean displacement derived from the recoilless effect is <x> ~1A°). The main idea of ref. 5 is that in effect, the dipolar field fluctuations are motionally averaged to zero, thus the line is emitted unbroadened. With harmonic fluctuations treated as frequency modulation of the line energy, the central line with natural width emerges exactly as in the recoil emission process itself. The side bands that create line broadening are resolved from the central line by $\xi = \Omega/\Gamma$ *line widths*[5]. Ω is the mean lattice vibrational (THz) frequency, thus $\xi = \Omega/\Gamma >>1$ since the nuclear width Γ is very small. This ensures complete isolation of the central line with the natural linewidth which explains the particular relevance to long lived states. Not only the dipolar fields, *all* line energy fluctuations originating from r–dependent interactions are also motionally narrowed via lattice vibrations. Thus a generalized *hypersharp* fraction is[5]:

$$\mathcal{H} = J_o^2\,(<x>/\lambda\,)\ \Pi_K\ J_o^2\,(\Delta_K\,/\hbar\,\Omega_K) \qquad (5),$$

where the first term is the familiar Debye-Waller factor -- the recoilless fraction-- and the Bessel functions in the product symbol represent the hypersharp fractions resulting from each of the other fluctuation. $\Delta_K$ is the maximum energy width of the fluctuation and $\Omega_K$, the rate, determined basically by the mean lattice vibration rate. In this generalized picture, recoilless emission is only one of many narrowing effects of lattice vibrations.

Other narrowed r dependent broadening are (see ref. 5 for details): the gravitational red shift ($\propto r$) averaged to zero and chemical shifts ($r^3$) and the zero point energy (ZPE $\propto r$) averaged to their *unique* equilibrium values so that the shifts are invariant from site to site. Similarly, external oscillations can be used to narrow the broadening via red shift distributions from practical size absorber/sources and possible differences in the earth's magnetic field at the source and absorber.

The only broadening not subject to motional narrowing is inhomogeneously distributed *static* multipole fields due to random lattice defects, thus it sets the basic limit on the observable linewidth. It would certainly negate hypersharp lines. In the $^3$H↔$^3$He case this is entirely *absent* because all moments higher than dipole are zero for the spin ½ $^3$H and $^3$He.

The T→He $\tilde{v}_e$ emission and absorption are precisely time reversed processes that ensure exact energy conservation. The $\tilde{v}_e$ energy is, in general, modified by energies $E_T$ and $E_{He}$ (which include shell electron binding energies, chemical shifts, the lattice vibration energy including the ZPE..). The second order Doppler effect (SOD) produces a shift $\Delta E/E = (3/2\ k\Delta T/\ Mc^2)$ that results in a net shift via the small difference in M (= 18.6 keV). The net SOD shift can be zeroed by canceling $\Delta T$ by identical (or the same), cryogenic baths for source/absorber. A fixed net shift due to the residual earth's field (after cancellation) may be unavoidable.

If the energy shifts $E_T$ and $E_{He}$ are *unique, static, and identical* in source/absorber, a deficit ($E_T$-$E_{He}$) in the BB decay T→He at $\tilde{v}_e$ emission is self-compensated *exactly* by ($E_{He}$-$E_T$) in the reverse He→T e-capture in $\tilde{v}_e$ absorption. An example of this effect is the role of the shell electron energy B in eq. 1 and 3.

Recoilless and hypersharp $\tilde{v}_e$ emission in TBB→$^3$He requires T and $^3$He (normally gases) to be embedded in solids. Metal tritides[7] offer a practical approach. Hydrogen (T) reacts with metals to form hydrides (tritides) and creates a uniform population of T in the bulk of the metal. As the tritide ages, the $^3$He daughter grows and populates the lattice ("the tritium-trick" TT). The He site in the source is its birth site –that of its parent T. The absorber is made in an identical manner. However, the absorber site of He, an insoluble mobile inert atom, is typically different and indeed, non-unique. He rapidly diffuses away and forms clusters/micro-bubbles, sites very different from regular lattice sites in T, thus basically unsuitable for hypersharp $\tilde{v}_e$ resonance. In bcc metals (Ti, Nb, V), the T sits only in tetragonal interstitial sites (TIS) whereas in fcc metals (Pd, Ni), it finds octahedral interstitial sites (OIS).

The key design problem is thus the search for a metal system where He sites are *lattice sites* identical to the T. A search was made using measured parameters from state of the art tritide research: 1) He diffusivity D(K) at temperature K, 2) the He generation rate g =(T/Metal M)x1.79x10⁻⁹/s, and 3) the activation energies E1, E2 and E3 for jumps, pair cluster formation and bubble coalescence[8] (see Table 1). A set of coupled non-linear differential equations[8] describe the time evolution of the concentrations c1 (mobile interstitials), c2 (pair clusters) and c3 (bubbles). These equations were solved numerically[9] for a variety of tritides, focusing on NbT and TiT. Fig. 1 shows the results for NbT. We see the growth of He for 200 days at which time the T is switched off by desorption. Thereafter, the He in the T-free sample has different ratios of (interstitial sites IS/(bubbles) = c1/(c2+c3) at different temperatures, exemplified by a flat c1 (as at 200K) or a decaying c1 indicating loss to growing bubbles c2+c3. The latter is shown explicitly in the lower curves for T>235K. The 200K results in Fig. 1 are not very sensitive to the exact parameter values: x100 larger D and/or a smaller E1 = 0.8 eV do not change the results. Thus, in NbT the $^3$He reside *only* at unique IS sites if T <200K, grows indefinitely and remains without bubble formation after the T is removed. This behavior in NbT is exceptional. In most other tritides e.g. in the well known PdT, bubble formation dominates already at >20K, normally leaving no He at regular lattice sites.

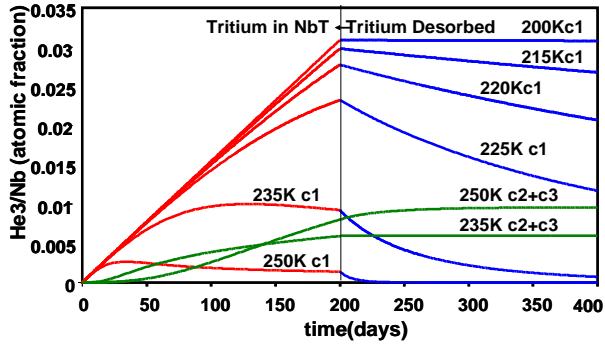

Fig 1 $^3$He is generated in NbT for 200 days, when the He generating T is desorbed. The figure shows the $^3$He concentration in interstitial sites (c1)(t) and that in clusters/ bubbles (c2+c3) (t) for different ambient temperatures and time t. The He in the T-free absorber below 220K is almost all interstitial and above 235K, almost all in clusters/ bubbles.

In the bcc NbT source, the T and thus, the just-born daughter He reside in the TIS[10] (see Table 2 for the Self-trapping energy EST at TIS and OIS). In the target the He would normally reside only in OIS (bcc or fcc) as e.g. in TiT. However, *Nb is unique* with degenerate EST(TIS) and EST(OIS) for He (Table 2). Thus, both sites can be randomly filled with equal probability. The site identifications have been verified by ion channeling.[11] There are 6 TIS and 3 OIS in the bcc unit cell. The ZPE at the two sites (Table 2) differ by ~0.01eV so that the 33% He in OIS are off resonant, a small cost. Thus, NbT aged below 200K offers unique, TIS He sites identical to that of the emitter T-He and (most of) the absorber He. In contrast, in TiT the emitter T is in a TIS but the absorber He sits *only* in the off-resonant OIS. The NbT system thus uniquely meets the stringent demands of a viable T-He matrix.

*Table 1  He transport parameters in NbT at 200K*

| M₁T₁ | E1 eV | E2 eV | E3 eV | D/cm² s |
|---|---|---|---|---|
| M=Nb | 0.9[a] | 0.13[b] | 0.43[b] | 1.1E-26[c] |

[a] Ref. 7; [b] Ref. 8; [c] Assumes *tritium* pre-exponential D₀ (ref. 7)

*Table 2. Theoretical lattice energy data for T and $^3$He in Nb interstitial sites (IS) (Ref. 10)*

| Site | EST (eV) | | ZPE (eV) | |
|---|---|---|---|---|
| | T | He | T | He |
| TIS | -0.133 | -0.906 | 0.071 | 0.093 |
| OIS | -0.113 | -0.903 | 0.063 | 0.082 |

The hypersharp fraction $\mathcal{H} \approx f$, basically because in the best present judgement, no significant source of non-harmonic fluctuations (that do not lead to motional narrowing) is evident. The fearsome inhomogeneous broadening from multiple static fields is absent in the T-He case. However, motion need not be exclusively harmonic. Stochastic motions such as sudden jumps in diffusion are known in tritides[12] at high temperatures and high concentrations. They are less applicable in the low concentrations here. These motions are not narrowed, they create kHz broadening. Other types of relaxation unrelated to the nuclear coordinate r, thus, to lattice vibrations, may also exist. The time scales of harmonic (THz) and stochastic (kHz) and other types of motions are so different that they do not interfere. The harmonic motional $\tilde{v}_e$ is emitted with hypersharp widths and cross sections. The $\tilde{v}_e$ from non-harmonic relaxation modes are broadened orders of magnitude wider with vastly smaller σ. Thus the effect is only equivalent to a $\tilde{v}_e$ flux loss which can be coped with by stronger sources/better geometry etc.

We can now sketch the bare experiment. We envision NbT sources and absorbers (gram scale to avoid *crystal recoil* to hypersharp precision!) made in an identical manner by the TT method of growing He by the decay of T. This implies a large T content in the absorber which will be a background (Tβ) against the signal of $\tilde{v}_e$ induced T activity (Rβ). A chief design goal is to achieve a large Rβ/Tβ. The source and absorber are set on a suitable baseline (~1cm in the initial phase) in the same cryogenic bath at temperatures << 200K. The signal is derived from the activation 18.6 keV betas (Rβ) from the absorber and the deviation caused by its growth from the known decay profile of T-betas (Tβ).

*Table 3 . T-³He hypersharp capture rates. Δt is the counting delay after start of activation. TT is the "tritium trick".*

| He abs. | Base line | T | ³He | Rβ/d | Tβ/d |
|---------|-----------|------|------|------------------------|----------|
| TT | 1 cm | 1kCi | 1 pg | $10^7$ (Δt=100d) | $3.7 \times 10^9$ |
| He Implant T-desorb | 1cm | 1mCi | 1 pg | 10 (Δt=0 d) | ~0 |
| He Implant T-desorb | 10m | 1kCi | 1 pg | 10 (Δt=0 d) | ~0 |

Using the ZPE (Z) data in Table 2 for He and T in Nb(TIS) the recoil free fraction f = f(He)f(T)≈ exp–{(27$E_R$/16) [1/Z(T)+1/Z(He)]}≈ 0.076. Table 3 shows signal rates in a primitive longitudinal geometry of source-absorber. The 1cm baseline (top line) can test most of the design optimization. The 1kCi source is state of the art[13]. The He absorber is made with the TT method and uses only a 1μCi T source to grow 1pg of He in 100 days. The signal rate grows linearly (a signature of $\bar{v}_e$ activation) and after a delay of 100 days, the signal Rβ is 100 Hz vs. the background Tβ of 37 kHz. The S/B is ~1/400, sufficient to confirm the signal growth vs. the T-decay.

For longer baselines, e.g. 10m needed to observe $\theta_{13}$ oscillations, the signal rate must be upped by $10^6$ mostly via the target mass. In the TT method this entails GHz Tβ rates in the absorber that present serious counting problems. Desorption of the T activity by ~$10^6$ via H exchange is considered practical[14]. An alternate He loading approach, low energy low temperature ion implantation of He in NbT[15], could be practical because of the very small doses of He involved. Both T desorption and implantation eliminate the deadweight of T activity in the absorber altogether. Applied to short baselines (line 1 in Table 3) this can similarly drastically reduce the source strength needed

The estimates of Table 3 show good prospects that hypersharp resonant $\bar{v}_e$ capture could be observed in the not too distant future, revolutionizing $v_e$ experiments. An important objective is oscillations via $\theta_{13}$ mixing, in the forefront now with large reactor based experiments on km-scale baselines (e.g., Dayabay, Double-Chooze). The low energy of the tritium $\bar{v}_e$ needs only 9.3m for equivalent sensitivity. The energy specificity of the signal makes it a text book test of $v_e$ oscillations[16]. Active-sterile mixing fits well in the program since tests for a sterile state with $\Delta m^2$ ~1eV² needs a baseline of only ~6 cm. The search for sterile states can thus be extended to a wide area of the $\Delta m^2$ parameter space.

Mead[17] suggested that a fundamental length (Planck length) $\mathcal{L}$ in nature would limit the ultimate widths of nuclear states i.e. ΔE/E depends on $\mathcal{L}$. With the definition $\mathcal{L}$= (G $\hbar$/c³)$^{1/2}$ ~$10^{-33}$ cm, Mead predicts ΔE/E($\mathcal{L}$) =$\mathcal{L}$ ($\mathcal{L}$/R) β~ $10^{-20}$ (for β = 1) to ΔE/E($\mathcal{L}$) ~$10^{-40}$ (for β=$\mathcal{L}$/R) (R is the nuclear radius). The form of β (in general ($\mathcal{L}$/R)$^n$), depends on the quantum gravity model. The hypersharp tritium $\bar{v}_e$ resonance offers the ideal probe of this prediction.. Indeed, a width ΔE/E~$10^{-20}$ implies a Planck broadening by $10^9$ of the tritium $\bar{v}_e$ resonance. A dilution of σ by this large a factor can be easily detected. Observation of the T $\bar{v}_e$ resonance would already preclude β = 1. Further results from scans of the resonance can influence quantum gravity models.

I wish to thank Don Cowgill and his colleagues at Sandia National Lab, Richard Kadell, Kung Hwa Park and many other colleagues for useful discussions, particularly on tritide technology and applications.

[1] J. N. Bahcall, Phys. Rev. **124** (1961) 495

[2] M.Visscher Phys. Rev. **116** (1959) 1581

[3] W. Kells and J .Schiffer, Phys. Rev. **C28** (1983) 2162

[4] R. S. Raghavan, ArXiv: hep-ph/06 1079 (2006)

[5] R. S. Raghavan, arXiv: 00805.4135 (2008)

[6] L. Mikaelyan et al, Sov J. Nucl. Phys **6** (1968) 254

[7] R. Lässer, *Tritium and ³He in Metals* (Springer, 1989)

[8] D. F. Cowgill, Sandia Natl. Lab. Report 2004-1739

[9] I thank Kunghwa Park and D. Rountree for their help in these calculations.

[10] M. J. Puska  R. M. Nieminen, Phys. Rev. **B29** (1984) 5382

[11] S. T. Picraux, Nucl. Inst. Meth. **182-183** (1981) 413; H. D.Carstanjen. Phys. Status. Solidi **A59** (1980) 11

[12] M. E. Stoll  T. J. Majors, Phys. Rev. **B24** (1981) 2859

[13] G. C. Abell & S. Attalla, Phys. Rev. Lett. **59** (1987) 995 used ~0.5 kCi T in 2.4 g PdH0.6 for low-T NMR studies.

[14] Demonstrated in G.C. Abell & D.F. Cowgill, Phys. Rev. **B44** (1991)4178; C. Weinheimer (Priv. Comm.)

[15] D. Cowgill, ref. 8 and private communication. This method, applied in many tritides, remains to be tested for NbT.

[16] E. Akhmedov et al, JHEP **0803** (2008) 005 (arXiv: 0802.2513 hep-ph) and arXiv 0803.1424 hep-ph  showed that tritium $\bar{v}_e$ of natural width oscillate, rebutting S. Bilenky et al (arxiv 0803. 0527-hep-ph) who claimed that they do not because of the time-energy uncertainty

[17] C. A. Mead, Phys. Rev. **143** (1965) 990